%% file: singleIas.tex
\documentclass[twocolumn]{aa}  
\usepackage[utf8]{inputenc}

\usepackage{amsmath}
\usepackage{nicefrac}

\usepackage{graphicx}
\usepackage{txfonts}
\usepackage{multirow}
\usepackage[colorlinks,citecolor=blue,urlcolor=blue]{hyperref}

\input{definitions.tex}

\graphicspath{{./}{figures/}}

\begin{document}

\title{Type Ia supernovae from non-accreting progenitors}

   \author{J. Antoniadis \inst{1,2}
   \and S. Chanlaridis \inst{2}
   \and G. Gr\"{a}fener\inst{2}
   \and N. Langer\inst{2,1}
          }

   \institute{Max Planck Institut f\"{u}r Radioastronomie
              Auf dem H\"{u}gel 69, 53121, Bonn, Germany
         \and
             Argelander Institut f\"{u}r Astronomie, Auf dem H\"{u}gel 71, 53121, Bonn, Germany \\
             \email{jantoniadis@mpifr-bonn.mpg.de}
             }


\abstract 
{Type Ia supernovae (\ias) are  manifestations of  stars deficient of hydrogen and helium disrupting in a 
thermonuclear runaway. While explosions of carbon-oxygen white dwarfs are 
thought to account for the majority of events, part of the observed diversity 
may be due to varied progenitor channels. We demonstrate that 
helium stars with masses between $\sim$1.8 and 2.5\msun\ may evolve 
into highly degenerate, near-Chandrasekhar mass cores with 
helium-free envelopes that subsequently  ignite carbon and oxygen explosively at densities $\sim(1.8-5.9)\times 10^{9}$\denu. This happens either due to core growth from 
shell burning (when the core has a hybrid CO/NeO composition), 
or following ignition of residual carbon triggered by  
exothermic electron captures on \iso{Mg}{24} (for a 
NeOMg-dominated  composition).
 We argue that the resulting 
thermonuclear runaways is likely to prevent core collapse, leading to the complete disruption of the star.
 The available nuclear energy at the onset of 
 explosive oxygen burning suffices to create 
 ejecta with a kinetic energy of $\sim$10$^{51}$\,erg, as in typical \ias. 
  Conversely, if these  runaways  result in partial disruptions, the corresponding transients would resemble  
 SN\,Iax events similar to SN\,2002cx. 
If helium stars in this mass range indeed explode as \ias, then the frequency of events would be comparable to the observed SN Ib/c rates, thereby sufficing to account for the majority of \ias in star-forming galaxies.}


\maketitle

\section{Introduction} \label{sec:intro}

 Despite their central role in Astrophysics and Cosmology, 
 the origin and physics of Type Ia supernovae (\ias)
remain uncertain  \citep[][]{Maoz:2013hna}. 
Typical \ia  luminosities ($\sim 10^{43}$\ergs) and ejecta velocities  
($ \sim 10^{4}$\kms), require \iso{Ni}{56}  
masses and kinetic energies of order 
$\sim 0.6$\msun\ and $\sim 10^{51}$\erg\ respectively. 
 These properties suggest that \ias are most likely 
 stars that disrupt in  thermonuclear explosions, 
 rather than core-collapse events. 
 Carbon/oxygen white dwarfs (CO WDs) 
 approaching the Chandrasekhar-mass limit (\mch)
 are the most promising progenitor systems, as they can
 produce explosions broadly consistent with observations
 \citep{Nomoto:1982zz,Wang:2012za,Churazov:2014bga}. 
 
Conventional stellar evolution channels, produce stable CO WDs with masses below $\sim 1.1$\msun. 
Consequently, matter accretion onto the WD is required to trigger an 
explosion, either via stable  transfer from a 
donor star (single-degenerate channels; SD), or in a 
 merger event (double-degenerate  and core-degenerate channels; DD and CD).  
Thus far, all of the  binary models    
encounter substantial difficulties in providing 
a self-consistent model for \ias \citep{Livio:2018rue,soker2019}. 
For instance, SD channels require considerable fine-tuning
of the mass accretion rate for
the WD to grow in mass. In addition, the interaction between  
the SN blast and the donor star or the circumbinary material, is expected to produce  signatures which 
are rarely or never seen, e.g. some contribution to the SN  luminosity 
at early times \citep{Kasen:2009si}, radio synchrotron emission 
\citep{Harris:2016hfr}, and H$\alpha$ emission due to unburned hydrogen. 
DD mergers on the other hand may produce a variety of outcomes, 
ranging from prompt 
explosions to long-lived remnants, or the delayed formation of a neutron star \citep{Livio:2018rue}. 
In addition, their overall contribution to the observed \ia rate may be too low \citep{vanKerkwijk:2010he,claeys2014a,Sato:2015spa}. 

Over the past 50 years, systematic studies of \ia explosions have revealed a large
diversity in their properties \citep{Taubenberger:2017hoo}. 
Notable outliers include luminous 
\citep[e.g. SN\,1991T;][]{filippenko1992} and ultra-luminous  
\citep[e.g. SNLS-03D3bb;][]{Howell:2006vn} SNe; SN\,1991bg-like transients which
are faint and fast evolving  \citep{ruiz-lapuente1993};  peculiar SN 2002cx-like SNe, 
 also known as SNe\,Iax, that may be pure thermonuclear deflagrations of helium-deficient compact objects \citep[e.g.][]{Li:2003wja,Branch:2004tq,Magee:2016vnu,Jha:2017gwq, Magee:2018aui}
and SN\,2012ca-like events, dubbed SNe Ia-CSM, for which 
there is evidence for interaction with a dense circum-stellar 
medium \citep{Bochenek:2017vok}. 
Even among ``normal'' \ias there is appreciable
scatter in rise times, maximum luminosities, ejecta velocities and spectral evolution   
\citep[][]{Livio:2018rue}. 
Finally, there seems to be a correlation with environment, 
as star-forming  galaxies typically  host more,  and brighter \ias \citep{Maoz:2013hna}.

While part of this diversity can be understood within the framework of SD and  DD/CD  
families, there may exist additional 
evolutionary pathways leading to \ias. 
It is know since long that intermediate-mass stars form
degenerate cores after helium burning which may ignite carbon
explosively if the core mass is allowed to grow due to shell buning
\citep{rose1969,Arnett:1969,wheeler1978}. This scenario became unlikely
when it was made clear that the loss of the hydrogen envelope during the
AGB evolution ends the core growth rather early, which can only
potentially be prevented in the most massive AGB stars \citep[see][and references therein]{Iben:1983ts}. 
In this case, however, the emerging supernova would be a H- and He-rich
“SNe 1.5”.
\cite{waldman2006a} and \cite{waldman2008} have considered the possible connection between ONe cores and \ias in more detail, 
demonstrating that some helium stars with \one cores can explode before any significant deleptonization due to electron captures on \iso{Ne}{20} occurs.
Their models  still retain a small He-rich envelope at the end and therefore, it is again unclear if such transients would appear as classical \ias. 

In spite of the aforementioned works, ONe cores are generally thought to produce massive WDs or core-collapse electron-capture supernovae \citep[ECSNe; e.g.,][]{nomoto1991,gutierrez1996,Takahashi:2013ena}. 
However, recent three-dimensional hydrodynamical 
simulations of oxygen deflagrations in ONe cores at 
central densities $\ge 10^{10}\denu$, {\it viz.} after the onset of electron-captures on \iso{Ne}{20}, 
suggest that a large fraction of the star ($\ge 1\msun$) may be 
ejected in a so-called thermonuclear ECSN, leaving 
behind only a small bound remnant \citep{Jones:2016asr,Jones:2018ule}.  

In light of these results, we revisit the work of \cite{waldman2006a} and \cite{waldman2008} using modern tools and updated input physics. We  demonstrate that a 
thermonuclear runaway leading to a \ia can indeed be initiated during the late evolution of a degenerate core of neon-oxygen 
(NeO) or carbon-neon-oxygen (CNeO) composition as it  approaches  \mch. We show that near-\mch~ \one cores originating from intermediate mass helium stars ($\sim 1.8-2.5$\msun) ---a common product of binary interactions--- can ignite their residual carbon and oxygen explosively at  
densities $\lesssim 6\times 10^{9}\denu$, 
before the onset of {\rm $^{20}$Ne(e$^-$,$\nu_e$)$^{20}$F} 
electron-capture reactions at $\sim 10^{10}\denu$ (Section\,\ref{sec:2}). 
In addition, during the final evolutionary stages, the envelope inflates significantly and is lost promptly  via winds or due to binary interaction. Consequently, the star is helium free when brought to thermonuclear explosion.   
The available nuclear energy at the time of central oxygen ignition suffices to unbind the star 
and to yield ejecta with kinetic energies comparable to what is expected for classical \ias (Section\,\ref{sec:3}).
 This mechanism  does not require accretion from the binary companion and 
 therefore may contribute significantly to the \ia  rate in young stellar populations (Section\,\ref{sec:4}).

\section{\one cores: formation and evolution}\label{sec:2}
\subsection{Overview}
Degenerate stellar cores of NeO composition form inside stars with ZAMS masses  
between $\sim$7 and 11\msun\ \citep[e.g.][]{poelarends2007,Poelarends:2007ip,Farmer:2015afs}. 
After core helium burning, such stars enter a super-asymptotic giant branch 
(SAGB) phase, characterized by a dense CO  
core and an  extended hydrogen envelope.  
As the core becomes increasingly more degenerate, it cools substantially 
due thermal neutrino emission. An important consequence is that the critical 
temperature for \iso{C}{12} ignition is first attained off-center, creating a 
turbulent flame that propagates inwards \citep{siess2006}.

Carbon burning in SAGB stars may be affected by complex mixing processes 
due to a combination of inverse composition gradients, overshooting, 
semi-convection and rotation. The penetration of 
Ne/O/Na/Mg ashes into unburned regions, may impact significantly the propagation of
the burning front. Mixing generally reduces the rate of thermonuclear reactions, leaving 
behind substantial amounts of residual carbon. In extreme cases, the flame can be 
quenched completely, resulting in a hybrid structure,  with a CO core, 
surrounded by a NeO mantle \citep{Denissenkov:2013qaa}. 

The subsequent evolution and final fate of such stars depends  critically on the competition 
between neutrino cooling due to the presence of \iso{Na}{23}\iso{Ne}{23} and 
\iso{Mg}{25}\iso{Na}{25} Urca pairs, and compressional heating due to core growth from the helium burning shell \citep{Schwab:2017epw}. 
SAGB stars are subject to significant dredge-up  and 
thermally unstable shell burning. 
These effects may impact substantially the ability of the core to grow in mass fast enough. 

However, thermal pulses and dredge-up episodes do not 
occur when the hydrogen envelope 
is lost, e.g. due to strong winds, or via  interactions   in a multiple system \citep{poelarends2007,Poelarends:2007ip,Woosley:2019sdf}.  
In such a case, helium shell burning is stable, allowing  
the core to approach the Chandrasekhar mass limit. 
In what follows, we build detailed numerical models to 
investigate the combined effects 
of residual unburned \iso{C}{12}, Urca cooling and constant mass growth from shell 
burning, in the late evolution of \one cores that originate from helium stars. 

\subsection{Numerical Calculations: Input Physics}\label{sec:2.1}
We use \mesa (version 10386) to follow the evolution of two helium-star models, \textsc{m1} and \textsc{m2}, with  masses of 2.5 and 1.8\msun\ respectively. 
The initial models have uniform compositions with $Y=0.98$ and $Z=0.02$ \citep[solar abundances are taken from ][]{grevesse1998}. We employ a nuclear network that considers 43  isotopes, from \iso{H}{1} to \iso{Ni}{58}. Reaction rates are based on the \texttt{JINA reaclib v2.0} compilation \citep{cyburt2010}. Electron screening factors and cooling rates from thermal neutrinos are evaluated as in \cite{Farmer:2015afs}, and references therein. 
Weak interaction rates are taken from \cite{Suzuki:2015iry}. 
Wind mass-loss rates are calculated using \mesa's \texttt{Dutch} compilation of recipes  \citep{Paxton:2013pj},  which is composition-independent. We use ``Type 2'' opacities during and after core helium burning.

Our baseline  model considers convection, thermohaline and semi-convectional mixing. Convective stability is 
evaluated using the Ledoux criterion. By default, \mesa uses  standard mixing-length theory  \citep[MTL;][]{cox1968} for convective mixing and energy transport. However, following carbon burning, both our models develop 
dynamically-unstable super-Eddington envelopes, causing numerical 
difficulties. For this reason, we decided to employ the ``enhanced'' MLT 
option available in \mesa \citep{Paxton:2013pj}, which artificially reduces 
the super-adiabatic gradient leading to  an enhanced convective energy 
transport efficiency.  This  allows us to follow the evolution of the core 
after carbon burning without interruptions.We further discuss this choice and its impact on the envelope evolution and the final mass in Section\,\ref{sec:5}. 
The MLT mixing length parameter is set to $a_{\rm MLT}=2.0$ for both models. For 
thermohaline mixing we employ 
the \cite{kippenhahn1980} 
 treatment, setting $D_{\text{TH}} = 1.0$ for the diffusion coefficient. Semi-convection is evaluated following \cite{langer1983}, adopting an efficiency parameter of $\alpha_{{\rm SEM}} = 1.0$. Our models do not employ the predictive convective boundary mixing approach described in \cite{Paxton:2017eie}. 

While \textsc{m1} does not consider the effects of convective overshoot ($f_{\rm ov}=0.0$), in \textsc{m2}, we set $f_{\rm ov}=0.014$ across all convective boundaries, including the base of the 
carbon-burning flame. While mixing at this interface is unlikely \citep{lecoanet2016}, 
we use this as a means to  quench the flame before it reaches the center.  
Other processes such as rotation and thermohaline mixing can lead to the 
same outcome for similar initial helium core masses \citep{Farmer:2015afs}.
The \mesa inlists are publicly available\footnote{\url{https://zenodo.org/record/3580243\#.XfjNjpNKjUI}}. 
An  extended grid exploring a broad range of initial masses, 
metallicities, overshooting and wind parameters will be presented in an 
accompanying paper \citep{chanlaridis2019}.

\subsection{Simulation results}
\begin{figure*}[htb!]
  \centering 
  \includegraphics[width=1.00\textwidth]{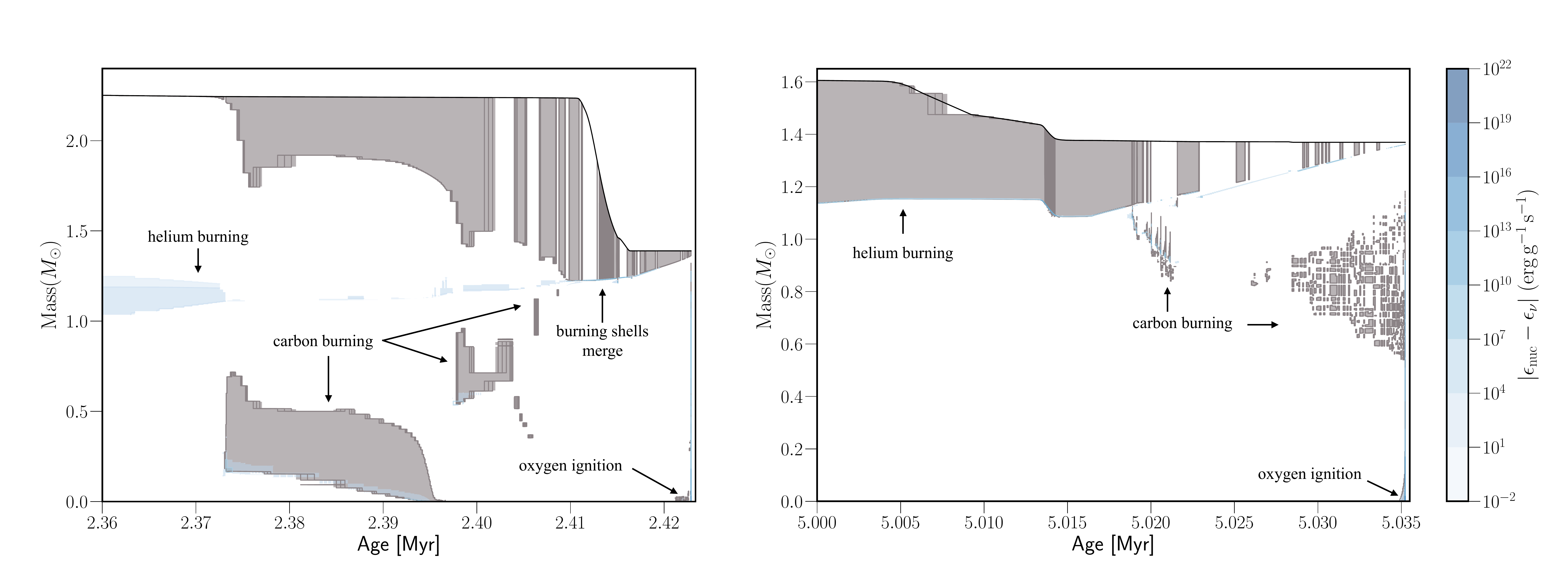}
  \caption{Kippenhahn diagrams following the evolution of \textsc{m1} (left) and \textsc{m2} after core helium depletion. Blue-shaded areas indicate regions in which nuclear burning occurs, i.e. locations for which the nuclear energy $\epsilon_{\rm nuc}$ exceeds energy losses due to neutrino emission, $\epsilon_\nu$. Gray regions are subject to convective mixing.$t=0$ corresponds to the onset of core helium burning.}
  \label{fig:1}
\end{figure*}

Figure\,\ref{fig:1} shows Kippenhahn diagrams for 
 \textsc{m1} and \textsc{m2}, focusing on the evolution after central helium depletion.
\textsc{m1} first ignites carbon at mass coordinate $\sim 0.3\msun$, when the total mass  is $ 2.25\msun$, and the CO core has a mass of $\sim 1.15\msun$.  
The initial flame is followed by secondary flashes propagating in both 
directions. Some of these episodes seem to occur only after a small critical 
mass of carbon has been accumulated below the burning shell. The entire carbon-burning phase lasts for about 
40,000\,yr. During most of this time, the star is a red giant with a low-density convective envelope  
($R\simeq 125\rsun$, $\logteff\simeq 3.75$, $\logl\simeq 
4.3$), and loses mass at a rate of  
$\mdot\simeq10^{-6}\mdotsun$, in good agreement with the
recent \textsc{kepler} models \citep{Woosley:2019sdf}, which employ the mass-loss scheme of \cite{Yoon:2017dme}, more specifically targeted to helium stars. While the mass-loss strength is highly uncertain, it has recently been suggested that the winds during the helium and carbon burning phases in low-mass helium stars may be less powerful than the \texttt{Dutch} schemes predict, given that Wolf-Rayet stars are not seen in this mass range \citep{Graefener2017,Vink:2017ujd}. In such a case, the cores in our models could grow faster.

As the core contracts and its surface gravity increases, the surrounding burning shells become progressively thinner. The envelope responds by expanding  and the stellar structure resembles closely that of an SAGB core. Interestingly, \cite{Woosley:2019sdf} finds that only helium stars with masses between $\sim 1.6$ and 3.2\msun, develop sufficiently thin helium-burning shells to cause envelope inflation. Hence, higher-mass stars would likely retain a significant fraction of outer layer until the end, leading to Type Ib SN explosions.

The last $\sim 5,000$\,yr of the evolution are characterized by vigorous burning in two neighbouring shells which eventually merge, resulting in violent burning, at $t\simeq 2.41$\,Myr (Figure\,\ref{fig:1}). In this phase, the star reaches extremely high luminosities up to $\log{L/L_\odot = 6.25}$, generating   a strong stellar wind that lasts for $\sim3000$\,yr and  eventually removes the remaining He-rich envelope.
The evolution of the envelope during this stage 
depends critically on the energy transport mechanism above Eddington lumonisities. With the enhanced MLT option 
employed in our calculations,  \textsc{m1} briefly 
becomes a yellow supergiant as the envelope  expands to $R\simeq 900\rsun$ while remaining dynamically stable.
The resulting strong wind of $\mdot\simeq10^{-3.8}\mdotsun$  is in the same range as theoretically expected maximum values for super-Eddington outflows \citep[][]{Owocki:2004zz,Smith2006}.
Conversely, using standard MLT, the envelope becomes dynamically unstable and our calculations encounter numerical difficulties just as the star leaves its Hayashi track, when the core has a mass of 1.32\msun.
By extrapolating the core-growth rate, the core of the star would likely still reach \mch. 
As our helium star may be the product of close binary evolution, at this stage further interactions would easily remove the envelope, as its binding energy corresponds to only a minuscule fraction of the orbital energy reservoir. 

Either way, the combination of enhanced mass loss  from both winds and binary interaction and vigorous burning, leads to the complete depletion of helium in the envelope. 
Following the neon flash, the small residual envelope contracts and the wind ceases completely for the last $\sim$5,000\,yr (Figure\,\ref{fig:1}). Our model stops when the star has a mass of 1.39\msun~ (see Sec.\,\ref{sec:runway}).  

The evolution of the envelope in \textsc{m2} is similar (Figure\,\ref{fig:1}). Here, 
the star expands twice, first for $\sim$5,000\,yr ($t=5.005$\,Myr in Figure\,\ref{fig:1}) and then again  briefly for some 
$\sim$500\,yr ($t=5.015$\,Myr), reaching a maximum size of $300\rsun$. The mass-loss rate mostly remains 
below $10^{-6}\mdotsun$. In \textsc{m2}, carbon ignites near mass coordinate $1.1\msun$, 
just as the star begins to develop an SAGB structure. The flame is quenched after only 
$0.1\msun$ of material has been converted to NeO, leaving behind a hybrid CO/NeO structure. Without overshooting, the flame propagates all the way to the center, converting the entire core to ONe (see model \textsc{m2}b in Figure\,\ref{fig:1}). The final mass of \textsc{m2} (and \textsc{m2}b) is 1.37\msun. 

To summarise, during the final evolutionary stages, both models are helium 
depleted and nearing \mch. The ability of the core 
to grow in mass depends somewhat on the uncertain mass-loss rate during the final burning phases. If the envelope 
is lost too early during the SAGB phase (which does not seem to be the case), then the two stars would 
leave behind white dwarfs with masses $\le 1.38\msun$, and ONe and CO/ONe composition respectively. 
Conversely, if the envelope is retained for long enough, as we find in our models, then the central density increases sufficiently to trigger either electron captures on \iso{Mg}{24} or central carbon ignition. 
In the following section we examine the evolution of the core during this phase.

\subsection{Oxygen ignition and thermonuclear runaway}\label{sec:runway}
\begin{figure*}[htb!]
\begin{center}
\includegraphics[width=1.00\textwidth]{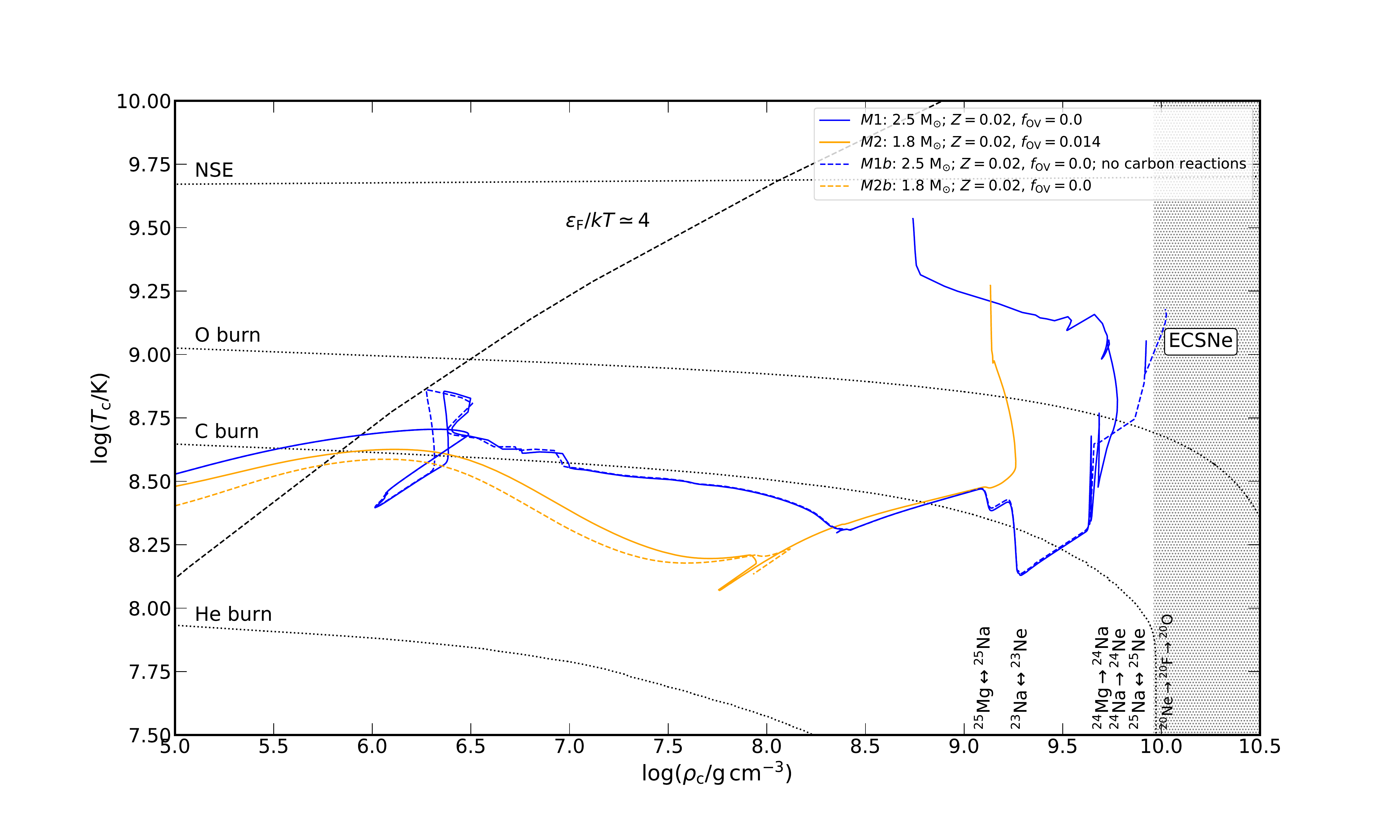}
\caption{Evolution of the core density and temperature for \textsc{m1} and \textsc{m2}. The dashed line shows the approximate boundary for electron degeneracy. Burning thresholds for a 100\% abundace of the corresponding species are indicated with dotted lines. The NSE threshold assumes an equilibrium timescale of 1\,s.  In model \textsc{m1}b, the energy contribution of carbon-consuming reactions is set to zero (see text). \textsc{M2}b shows the core evolution of a 1.8\msun without overshooting. While this model stopped at $\logrhoc \simeq 8.2$ due to numerical convergence issues, the final core mass is similar to \textsc{m2}.}
\label{fig:2}
\end{center}
\end{figure*}

Figure\,\ref{fig:2} gives an overview of the central density and temperature evolution for models  \textsc{m1} and \textsc{m2}.
Following the main carbon-burning episode, both stars continue to contract, while cooling  due to neutrino  emission. As shell 
burning intensifies, compressional heating eventually balances 
off neutrino losses, at $\logrhoc \simeq 8.3$ and 8.0 for \textsc{m1} and \textsc{m2} respectively. The subsequent evolution depends on the composition. 

For \textsc{m1}, the degenerate core is composed 
mostly of neon and oxygen. The most abundant isotopes have  $\abun{O}{16}\simeq 0.43$;  $\abun{Ne}{20}\simeq 0.42$; $\abun{Mg}{24}\simeq0.1$;  $\abun{C}{12}\simeq 0.011$; $\abun{Na}{23}\simeq 0.037$; $\abun{Mg}{25}\simeq 0.001$. Between $\logrhoc=9.05$ and 9.25, the temperature drops to $\logtc\simeq 8.2$ due to \iso{Mg}{25}\iso{Na}{25} and \iso{Na}{23}\iso{Na}{23} direct Urca reactions. At higher densities, neutrino cooling ceases completely, and the temperature rises again (Figure\,\ref{fig:2}). 
When $\logrhoc=9.65$, exothermic electron captures on \iso{Mg}{24} and 
\iso{Na}{24} start occurring at a substantial rate, raising the temperature 
adequately to ignite carbon. In turn, this triggers oxygen burning and a thermonuclear runaway at  $\logrhoc= 9.77$. This ignition density is  
lower than the   $\logrhoc\ge 9.97$ typically expected for oxygen deflagrations  in pure NeO cores \citep{Jones:2018ule}. As a consequence, 
severe deleptonization due to \iso{Ne}{20} electron captures is likely avoided.  To demonstrate the importance of residual carbon, we 
computed a variation of this model in which the energy contribution of all carbon-consuming reactions is artificially suppressed following the 
main carbon-burning phase (model \textsc{m1b} in Figure\,\ref{fig:2}). In this case, oxygen ignites above $\logrhoc\simeq 10$ and only after \iso{Ne}{20} electron captures have started to occur at a significant rate. Hence, in the absence of residual carbon, this star would produce a core-collapse or a thermonuclear ECSN \citep{Jones:2016asr}.

\textsc{m2} is composed mostly of carbon and oxygen, with $\abun{C}{12}=0.38$ and $\abun{O}{16}=0.60$ respectively. 
 Here, \iso{Na}{23} is not abundant enough to cause  substantial cooling. Consequently, carbon, which is significantly more abundant compared to \textsc{m1}, ignites at $\logrhoc=9.26$. 
 
The evolution following central oxygen ignition is not adequately modeled in our 1D simulations. 
 \textsc{m2} will most likely disrupt in a \ia, as the composition and
 ignition conditions resemble closely those found in standard CO WD \ia
 progenitors \citep{Nomoto:1982zz}. While the fate of \textsc{m1} is less 
 certain, a thermonuclear explosion is also the most likely outcome: 
 firstly, the available nuclear energy is sufficient to unbind the star (see 
 below). Secondly, the ignition density is only slightly higher than  what is typically considered for CO WD \ia progenitors. 
 Hence, the deflagration ashes will likely be buoyant, leading to expansion, which will in turn limit the deleptonization rate. This hypothesis is strongly supported
 by 3D hydrodynamic  simulations of ECSN deflagrations by \cite{Jones:2016asr,Jones:2018ule}: their least compact progenitor ingites at $\logrhoc=9.90$ but still manages to eject  $\sim 1\msun$ of material. 
 Similarly,  \cite{marquardt2015} simulate ONe WD detonations at lower 
 densities and demonstrate that the explosion is practically identical to a 
 typical \ia. 
 Interestingly in our 1D simulations, both models experience significant expansion. This is most likely the result of (over-)efficient convection, which also homogenizes the inner $\sim 1\msun$ of the core.

\subsection{Energetics and nucleosynthesis}\label{sec:3}
\begin{figure}
\begin{center}
\includegraphics[width=0.5\textwidth]{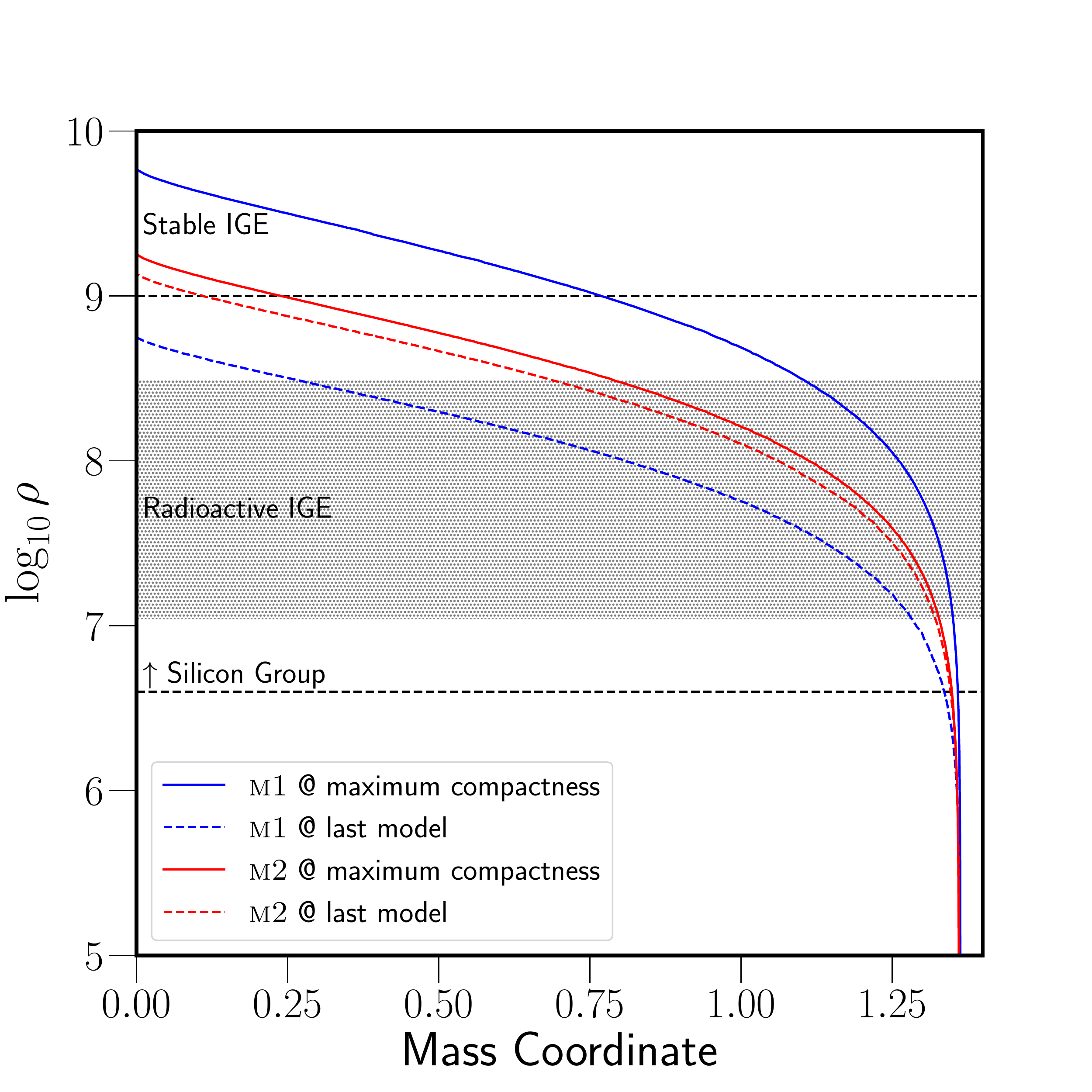}
\caption{Density profiles at maximum compactness (solid lines) and at the end of our \mesa calculations (dashed lines). Regions indicating approximate burning regimes as in \cite{Seitenzahl2017}.}
\label{fig:nuc}
\end{center}
\end{figure}

Figure\,\ref{fig:nuc} shows the density profiles of \textsc{m1} and \textsc{m2} at maximum compactness ($\logrhoc=9.77$ and 9.22, respectively), and at the end of our simulations. At the onset of oxygen ignition, \textsc{m1} and \textsc{m2} have binding energies of $E_{\rm bind}=-5.76\times10^{50}$ and $-5.16\times10^{50}$\,erg, and average electron fractions of $\ye= 0.496$ and  0.499, respectively. If these progenitors were to produce an \ia of  typical composition, with $\sim 0.7\msun$ of nickel and iron and 0.6\msun\ of Si-group elements, the corresponding kinetic energies of the ejecta would be $E_{\rm kin}=E_{\rm nuc}-E_{\rm bind}\simeq 0.83\times10^{51}$ and  $1.17\times10^{51}$\,erg. 
Obviously, the nucleosynthesis yields depend on the actual
\ye\ and density profiles during explosive burning. If 
\textsc{m1} achieves nuclear statistical equilibrium (NSE)
before any significant expansion and deleptonization, then it
would produce mostly stable iron-peak elements and $\sim 0.3\msun$ of \iso{Ni}{56}, resulting in a sub-luminous 
explosion that would resemble a SN\,Iax.  Similarly low nickel masses would be expected if the deflagration wave never transitions to a detonation. On the other hand, if NSE is achieved at densities similar to our last \mesa model, following an 
initial expansion (Figure\,\ref{fig:nuc}), more than 
1\msun\ of \iso{Ni}{56} can be produced, with only moderate
amounts of iron. Similarly, \textsc{m2} could produce up 
to 1.3\msun\ of iron elements if it does not expand any further. 

\section{Rates and delay times}\label{sec:4}
An additional  proxy for the potential connection between exploding \one cores and \ias, is a comparison between their  occurrence rates. 

A zeroth-order upper bound on the number of stripped near-Chandrasekhar mass \one cores per unit solar mass of star formation, $n_\star$, is 
simply the number of stars, $n_{\rm{He}1.8-2.5}$, that end up developing  helium cores in the relevant mass range, here taken to be $1.8\lesssim M_{\rm{He}}/\msun \lesssim 2.5$, based on our simulations and the recent results of \cite{Woosley:2019sdf} and \cite{Farmer:2015afs} who also find that carbon ignites off-center in this mass range, independently of convective overshoot assumptions.

 Based on 
\cite{Farmer:2015afs},  such helium stars originate from ZAMS stars with masses~$\sim$7$\ldots 11\msun$. Adopting a \cite{Chabrier:2004vw} initial mass function (IMF), for the aforementioned mass range one  finds,  $n_\star=n_{\rm{He}1.8-2.5}\simeq 0.0059$, which is larger than the observed number of \ias per solar mass, n$_{\rm \ia}\simeq 0.001-0.003$ \citep[e.g.][]{claeys2014a,Maoz:2013hna}. 
Here, one should keep in mind that the mass width for off-centre carbon burning is subject to several  physical uncertainties, e.g. 
related to winds, as well as mixing and flame propagation during the carbon burning phase (Section\,\ref{sec:2}). In addition, for isolated 
H-rich SAGB stars, the second  dredge-up also limits the mass range 
for which the helium core mass remains near \mch, to about $\sim0.1\msun$ around a ZAMS mass of 11\msun\ \citep[c.f. Figure\,1 in][]{Podsiadlowski:2003py}. Hence, more realistically, one would expect a maximum single-star contribution $\sim10$ times smaller than the rate estimated above. However, stars in interacting multiple systems, which loose their 
H-rich envelopes early, avoid dredge-up. Hence, most interacting stars capable of igniting carbon off-centre will also 
manage to reach near-\mch final core masses \citep[][and references therein]{Podsiadlowski:2003py,poelarends2007,Poelarends:2007ip,chen2014b,Meng:2014qta,doherty2015,Poelarends:2017dua}.

If this channel is indeed mostly relevant to stars in multiple systems, {\it viz.} the hydrogen envelope can only be removed via binary 
interactions, then only the fraction of $n_{\rm{He}1.8-2.5}$ in interacting binaries is relevant. 
Adopting the same assumptions for the initial mass range as above, $f_{\rm bin}=0.7$ for the fraction of stars in interacting binaries \citep{Sana:2012px}, a \cite{Chabrier:2004vw} IMF for the primaries (i.e. the 
initially more massive stars), and a mass-ratio distribution scaling as $\propto q^{-0.1}$ \citep{Sana:2012px}, one finds $n_\star \simeq 0.0038$. Here, systems for which both stars have initial masses between 7 and 11\msun\ are only counted once, as the secondary would evolve in isolation once the primary explodes. Owing to the shape of the IMF, about 80\% of the stars contributing to the former estimate are primaries, i.e. they would explode while still having a less-evolved companion. 

 More realistically, a significant number of these binaries would likely interact before the carbon-burning phase, thereby 
 affecting the growth of the core and consequently the event rate. While detailed  calculations are required to probe the 
 influence  of binary interactions, an upper limit on the expected rate can be obtained by comparing  the frequency of 
 progenitor systems  relative to  the SN Ib/c progenitors, assuming that the latter are dominated by stripped helium stars 
 with masses $\ge 2.5\msun$ \citep{Woosley:2019sdf}, which evolve similarly. Adopting the same IMF and initial mass range 
 as above, one finds a total rate of $\sim 0.8-1.0$ times the SN Ib/c rate, which is consistent with observations \citep[][and references therein]{branch2017}.

Besides the total number of \ias, a property that may be more challenging to match with observations is the evolution of 
the \ia rate with cosmic time. \ias from such a channel would have delay times dominated by the main sequence lifetime of the progenitors, i.e. of order 30 to 78\,Myr, for stars with ZAMS masses between 7 and 11\msun. Binaries interacting via early Case\,A RLO prior to the removal of the hydrogen envelope, could contribute events up to $\sim 300$\,Myr following star formation (corresponding to the MS lifetime of a 3.5\msun\ star). 
These delay times could help account for the high \ia rates in star-forming galaxies, compared to ellipticals \citep{Maoz:2010pz,claeys2014}, but some variant of the DD channel, or more generally a model in which the delay time much larger than the MS lifetime of intermediate-mass stars, would still be required to explain events with much longer delay times.  
 However,  if the \one core of a WD  were to evolve in a similar way due to 
mass accretion from a companion as in the SD scenarios, or in a merger similar to the DD and CD scenarios, then the corresponding explosion could contribute a \ia several Gyrs after star formation  \citep[see also][and references therein]{Kashi:2011nm,chen2014b,Meng:2014qta,marquardt2015,Schwab:2015bma,Jones:2016asr,Schwab:2017epw,Schwab:2018cnb,kashyap2018,augustine2019,soker2019}. 
Finally, if \one explosions produce less nickel than typical \ias, then the short delay times would match those of under-luminous SNe\,Iax which predominately occur in star-forming regions \citep{Lyman:2013drq,Jha:2017gwq}.

\section{Summary}\label{sec:5}
We have shown that at least some stars capable of developing near-\mch~ \one cores 
after losing their hydrogen  envelopes, may explode as \ias instead of 
undergoing a core-collapse  ECSN. For NeO compositions, the runaway seems to be triggered by the ignition of 
residual carbon following electron captures on \iso{Mg}{24}, which in turn 
leads to explosive oxygen burning at densities below $5.9\times 10^{9}\denu$ (Sections\,\ref{sec:2}). 
For hybrid CO/NeO cores, ignition is triggered by core compression, at a density of $1.8\times10^{9}\denu$, similar to what is expected for CO WD deflagrations. For either case, the conditions at the onset of oxygen burning are such that the energetics  could resemble closely a typical \ia, or an under-luminous Iax depending on when and if the initial deflagration transitions into a detonation (Section\,\ref{sec:3}). 
It would be worth considering whether the differences in initial density and 
composition could lead to distinct nucleosynthetic signatures that would help 
distinguish these progenitors and/or contribute uniquely to the chemical 
evolution of the Galaxy \citep[in analogy to ][for ECSNe]{Jones:2018ule}.

The frequency of the corresponding progenitor systems 
is sufficient to account for a considerable fraction of the observed \ia  rate.  Our optimistic zeroth-order upper bounds in  Section\,\ref{sec:4} suggest that \ias from non-accreting progenitors could occur at the same rate as SNe Ib/c, assuming that the latter also result from stripped helium stars. Since, the bulk of events would occur only $\sim 50$\,Myr after star formation, this channel is mostly relevant to star-forming galaxies. The shorter delay times compared to traditional \ia scenarios, open further interesting avenues for constraining this model, e.g.  by considering the abundance rations of alpha elements relative to iron in metal-poor stars. 

 While here we demonstrate that the helium envelope is  likely lost only  after the core has grown to \mch, its evolution remains a major uncertainty for this progenitor channel. If helium is removed sufficiently early, e.g. due to binary interactions, then sub-\mch~\one\ WDs would be formed instead. 
 
In spite of the mass-loss 
uncertainties, if viable, this channel would help explain some of the observed \ia diversity. Since either star in a binary system may potentially explode 
as a \ia without accreting from its  companion, the resulting events can resemble setups expected in all SD and DD/CD scenarios. Explosions of 
secondaries, would follow a first core-collapse SN. This 
could lead to \ia remnants with no luminous surviving stars, high proper motions due to a kick from the first SN \cite[like the Kepler SN remnant,][]{Chiotellis:2011jy}, and  possibly associated with a neutron star. 
Since the  envelope can be removed either due to winds, case-BB mass transfer, or a common envelope event, some diversity is also expected in 
the SN environment. In turn this would influence both the 
appearance of the explosion and the evolution of the SN remnant. The rates estimated in Section\,\ref{sec:4} are also broadly consistent with the 
number of \ias that seem to explode inside planetary 
nebulae \citep[$\sim 20\%$;][]{Tsebrenko:2014kea}.

\begin{acknowledgements}
We thank the referee for the extremely helpful report. 
 This research made extensive use of NASA's ADS, \textsc{mesa}\footnote{http://mesastar.org} \citep{Paxton:2010ji,Paxton:2013pj,Paxton:2015jva,Paxton:2017eie} and Astropy\footnote{http://www.astropy.org} \citep{Price-Whelan:2018hus}. Figure 1 was created with NuGridPy\footnote{https://github.com/NuGrid/NuGridPy}
\end{acknowledgements}

\bibliography{singleIas} 
\bibliographystyle{yahapj}

\end{document}

%% file: definitions.tex
\usepackage{xspace} 

\newcommand{\iso}[2]{\ensuremath{\mathrm{{^{#2}}#1}}} 
\newcommand{\abun}[2]{\ensuremath{X(\mathrm{{^{#2}}#1}})} 

\newcommand{\msun}{\ensuremath{\mathrm{\,M_\odot}}} 
\newcommand{\rsun}{\ensuremath{\mathrm{\,R_\odot}}} 
\newcommand{\lsun}{\ensuremath{\mathrm{\,L_\odot}}} 
\newcommand{\mdotsun}{\ensuremath{\mathrm{\,M_{\odot}\,yr^{-1}}}} 
\newcommand{\erg}{\,\ensuremath{\mathrm{erg}}} 
\newcommand{\ergs}{\ensuremath{\mathrm{\,erg\,s^{-1}}}} 
\newcommand{\kms}{\ensuremath{\mathrm{\,km\,s^{-1}}}} 
\newcommand{\denu}{\ensuremath{\mathrm{\,g\,cm^{-3}}}} 

\newcommand{\logtc}{\ensuremath{\log_{10} (T_{\mathrm{c}}/{\mathrm{K})}}} 
\newcommand{\logl}{\ensuremath{\log_{10} (L/\lsun)}} 
\newcommand{\logrhoc}{\ensuremath{\log_{10} (\rho_{\mathrm{c}}/ \mathrm{g\,cm^{-3}})}} 
\newcommand{\logteff}{\ensuremath{\log_{10} (T_{\mathrm{eff}} /\mathrm{K})}} 
\newcommand{\mch}{\ensuremath{M_{\mathrm{Ch}}}} 
\newcommand{\mdot}{\ensuremath{\dot{M}}} 
\newcommand{\ye}{\ensuremath{Y_e}} 

\newcommand{\mesa}{{\textsc{mesa}}\xspace} 
\newcommand{\ia}{SN\,Ia\xspace} 
\newcommand{\ias}{SNe\,Ia\xspace} 
\newcommand\one{(C)N\lowercase{e}O\xspace} 

%% file: singleIas.bbl
\begin{thebibliography}{77}
\providecommand\natexlab[1]{#1}
\providecommand\JournalTitle[1]{#1}

\bibitem[{Arnett(1969)}]{Arnett:1969}
Arnett, W.~D. 1969,
  \href{http://dx.doi.org/10.1007/BF00650291}{\JournalTitle{The Astrophysical
  Journal Supplement Series}, 5, 180}

\bibitem[{{Astropy Collaboration} \& {Astropy
  Contributors}(2018)}]{Price-Whelan:2018hus}
{Astropy Collaboration}, \& {Astropy Contributors}. 2018,
  \href{http://dx.doi.org/10.3847/1538-3881/aabc4f}{\JournalTitle{The
  Astronomical Journal}, 156, 123}

\bibitem[{Augustine {et~al.}(2019)Augustine, Willcox, Brooks, Townsley, \&
  Calder}]{augustine2019}
Augustine, C.~N., Willcox, D.~E., Brooks, J., Townsley, D.~M., \& Calder, A.~C.
  2019, \JournalTitle{arXiv e-prints}, 1910, arXiv:1910.12403

\bibitem[{Bochenek {et~al.}(2018)Bochenek, Dwarkadas, Silverman, Fox,
  Chevalier, Smith, \& Filippenko}]{Bochenek:2017vok}
Bochenek, C.~D., Dwarkadas, V.~V., Silverman, J.~M., {et~al.} 2018,
  \href{http://dx.doi.org/10.1093/mnras/stx2029}{\JournalTitle{Monthly Notices
  of the Royal Astronomical Society}, 473, 336}

\bibitem[{Branch {et~al.}(2004)Branch, Baron, Thomas, Kasen, Li, \&
  Filippenko}]{Branch:2004tq}
Branch, D., Baron, E., Thomas, R.~C., {et~al.} 2004,
  \href{http://dx.doi.org/10.1086/425081}{\JournalTitle{Publications of the
  Astronomical Society of the Pacific}, 116, 903}

\bibitem[{Branch \& Wheeler(2017)}]{branch2017}
Branch, D., \& Wheeler, J.~C. 2017,
  \href{http://dx.doi.org/10.1007/978-3-662-55054-0}{\JournalTitle{Supernova
  Explosions: Astronomy and Astrophysics Library, ISBN 978-3-662-55052-6.
  Springer-Verlag GmbH Germany, 2017}}

\bibitem[{Chabrier(2005)}]{Chabrier:2004vw}
Chabrier, G. 2005,
  \href{http://dx.doi.org/10.1007/978-1-4020-3407-7_5}{\JournalTitle{The
  Initial Mass Function 50 Years Later}, 327, 41}

\bibitem[{Chanlaridis {et~al.}(2019)Chanlaridis, Antoniadis, Gr\"{a}fener, \&
  Langer}]{chanlaridis2019}
Chanlaridis, S., Antoniadis, J., Gr\"{a}fener, G., \& Langer, N. 2019,
  \JournalTitle{in prep.}

\bibitem[{Chen {et~al.}(2014)Chen, Herwig, Denissenkov, \& Paxton}]{chen2014b}
Chen, M.~C., Herwig, F., Denissenkov, P.~A., \& Paxton, B. 2014,
  \href{http://dx.doi.org/10.1093/mnras/stu108}{\JournalTitle{Monthly Notices
  of the Royal Astronomical Society}, 440, 1274}

\bibitem[{Chiotellis {et~al.}(2012)Chiotellis, Schure, \&
  Vink}]{Chiotellis:2011jy}
Chiotellis, A., Schure, K.~M., \& Vink, J. 2012,
  \href{http://dx.doi.org/10.1051/0004-6361/201014754}{\JournalTitle{Astronomy
  \& Astrophysics}, 537, A139}

\bibitem[{Churazov {et~al.}(2014)Churazov, Sunyaev, Isern, Kn{\"o}dlseder,
  Jean, Lebrun, Chugai, Grebenev, Bravo, Sazonov, \& Renaud}]{Churazov:2014bga}
Churazov, E., Sunyaev, R., Isern, J., {et~al.} 2014,
  \href{http://dx.doi.org/10.1038/nature13672}{\JournalTitle{Nature}, 512, 406}

\bibitem[{Claeys {et~al.}(2014{\natexlab{a}})Claeys, Pols, Izzard, Vink, \&
  Verbunt}]{claeys2014a}
Claeys, J. S.~W., Pols, O.~R., Izzard, R.~G., Vink, J., \& Verbunt, F. W.~M.
  2014{\natexlab{a}},
  \href{http://dx.doi.org/10.1051/0004-6361/201322714}{\JournalTitle{Astronomy
  \& Astrophysics}, 563, A83}

\bibitem[{Claeys {et~al.}(2014{\natexlab{b}})Claeys, Pols, Izzard, Vink, \&
  Verbunt}]{claeys2014}
---. 2014{\natexlab{b}},
  \href{http://dx.doi.org/10.1051/0004-6361/201322714}{\JournalTitle{Astronomy
  and Astrophysics}, 563, A83}

\bibitem[{{Cox} \& {Giuli}(1968)}]{cox1968}
{Cox}, J.~P., \& {Giuli}, R.~T. 1968, {Principles of stellar structure}

\bibitem[{Cyburt {et~al.}(2010)Cyburt, Amthor, Ferguson, Meisel, Smith, Warren,
  Heger, Hoffman, Rauscher, Sakharuk, Schatz, Thielemann, \&
  Wiescher}]{cyburt2010}
Cyburt, R.~H., Amthor, A.~M., Ferguson, R., {et~al.} 2010,
  \href{http://dx.doi.org/10.1088/0067-0049/189/1/240}{\JournalTitle{The
  Astrophysical Journal Supplement Series}, 189, 240}

\bibitem[{Denissenkov {et~al.}(2013)Denissenkov, Herwig, Truran, \&
  Paxton}]{Denissenkov:2013qaa}
Denissenkov, P.~A., Herwig, F., Truran, J.~W., \& Paxton, B. 2013,
  \href{http://dx.doi.org/10.1088/0004-637X/772/1/37}{\JournalTitle{The
  Astrophysical Journal}, 772, 37}

\bibitem[{Doherty {et~al.}(2015)Doherty, {Gil-Pons}, Siess, Lattanzio, \&
  Lau}]{doherty2015}
Doherty, C.~L., {Gil-Pons}, P., Siess, L., Lattanzio, J.~C., \& Lau, H. H.~B.
  2015, \href{http://dx.doi.org/10.1093/mnras/stu2180}{\JournalTitle{Monthly
  Notices of the Royal Astronomical Society}, 446, 2599}

\bibitem[{Farmer {et~al.}(2015)Farmer, Fields, \& Timmes}]{Farmer:2015afs}
Farmer, R., Fields, C.~E., \& Timmes, F.~X. 2015,
  \href{http://dx.doi.org/10.1088/0004-637X/807/2/184}{\JournalTitle{The
  Astrophysical Journal}, 807, 184}

\bibitem[{Filippenko {et~al.}(1992)Filippenko, Richmond, Matheson, Shields,
  Burbidge, Cohen, Dickinson, Malkan, Nelson, Pietz, Schlegel, Schmeer,
  Spinrad, Steidel, Tran, \& Wren}]{filippenko1992}
Filippenko, A.~V., Richmond, M.~W., Matheson, T., {et~al.} 1992,
  \JournalTitle{The Astrophysical Journal}, L5

\bibitem[{{Gr{\"a}fener} {et~al.}(2017){Gr{\"a}fener}, {Owocki}, {Grassitelli},
  \& {Langer}}]{Graefener2017}
{Gr{\"a}fener}, G., {Owocki}, S.~P., {Grassitelli}, L., \& {Langer}, N. 2017,
  \href{http://dx.doi.org/10.1051/0004-6361/201731590}{\JournalTitle{\aap},
  608, A34}

\bibitem[{Grevesse \& Sauval(1998)}]{grevesse1998}
Grevesse, N., \& Sauval, A.~J. 1998,
  \href{http://dx.doi.org/10.1023/A:1005161325181}{\JournalTitle{Space Science
  Reviews}, 85, 161}

\bibitem[{Gutierrez {et~al.}(1996)Gutierrez, {Garcia-Berro}, Iben, Isern,
  Labay, \& Canal}]{gutierrez1996}
Gutierrez, J., {Garcia-Berro}, E., Iben, I., {et~al.} 1996,
  \href{http://dx.doi.org/10.1086/176934}{\JournalTitle{The Astrophysical
  Journal}, 459, 701}

\bibitem[{Harris {et~al.}(2016)Harris, Nugent, \& Kasen}]{Harris:2016hfr}
Harris, C.~E., Nugent, P.~E., \& Kasen, D.~N. 2016,
  \href{http://dx.doi.org/10.3847/0004-637X/823/2/100}{\JournalTitle{The
  Astrophysical Journal}, 823, 100}

\bibitem[{Howell {et~al.}(2006)Howell, Sullivan, Nugent, Ellis, Conley,
  Le~Borgne, Carlberg, Guy, Balam, Basa, Fouchez, Hook, Hsiao, Neill, Pain,
  Perrett, \& Pritchet}]{Howell:2006vn}
Howell, D.~A., Sullivan, M., Nugent, P.~E., {et~al.} 2006,
  \href{http://dx.doi.org/10.1038/nature05103}{\JournalTitle{Nature}, 443, 308}

\bibitem[{Iben \& Renzini(1983)}]{Iben:1983ts}
Iben, Jr., I., \& Renzini, A. 1983,
  \href{http://dx.doi.org/10.1146/annurev.aa.21.090183.001415}{\JournalTitle{Annual
  Review of Astronomy and Astrophysics}, 21, 271}

\bibitem[{Jha(2017)}]{Jha:2017gwq}
Jha, S.~W. 2017,
  \href{http://dx.doi.org/10.1007/978-3-319-21846-5_42}{\JournalTitle{Handbook
  of Supernovae}, 375}

\bibitem[{Jones {et~al.}(2016)Jones, R{\"o}pke, Pakmor, Seitenzahl, Ohlmann, \&
  Edelmann}]{Jones:2016asr}
Jones, S., R{\"o}pke, F.~K., Pakmor, R., {et~al.} 2016,
  \href{http://dx.doi.org/10.1051/0004-6361/201628321}{\JournalTitle{Astronomy
  \& Astrophysics}, 593, A72}

\bibitem[{Jones {et~al.}(2019)Jones, R{\"o}pke, Fryer, Ruiter, Seitenzahl,
  Nittler, Ohlmann, Reifarth, Pignatari, \& Belczynski}]{Jones:2018ule}
Jones, S., R{\"o}pke, F.~K., Fryer, C., {et~al.} 2019,
  \href{http://dx.doi.org/10.1051/0004-6361/201834381}{\JournalTitle{Astronomy
  and Astrophysics}, 622, A74}

\bibitem[{Kasen(2010)}]{Kasen:2009si}
Kasen, D. 2010,
  \href{http://dx.doi.org/10.1088/0004-637X/708/2/1025}{\JournalTitle{The
  Astrophysical Journal}, 708, 1025}

\bibitem[{Kashi \& Soker(2011)}]{Kashi:2011nm}
Kashi, A., \& Soker, N. 2011,
  \href{http://dx.doi.org/10.1111/j.1365-2966.2011.19361.x}{\JournalTitle{Monthly
  Notices of the Royal Astronomical Society}, 417, 1466}

\bibitem[{Kashyap {et~al.}(2018)Kashyap, Haque, {Lor{\'e}n-Aguilar},
  {Garc{\'i}a-Berro}, \& Fisher}]{kashyap2018}
Kashyap, R., Haque, T., {Lor{\'e}n-Aguilar}, P., {Garc{\'i}a-Berro}, E., \&
  Fisher, R. 2018,
  \href{http://dx.doi.org/10.3847/1538-4357/aaedb7}{\JournalTitle{The
  Astrophysical Journal}, 869, 140}

\bibitem[{Kippenhahn {et~al.}(1980)Kippenhahn, Ruschenplatt, \&
  Thomas}]{kippenhahn1980}
Kippenhahn, R., Ruschenplatt, G., \& Thomas, H.-C. 1980,
  \JournalTitle{Astronomy and Astrophysics}, 91, 175

\bibitem[{Langer {et~al.}(1983)Langer, Fricke, \& Sugimoto}]{langer1983}
Langer, N., Fricke, K.~J., \& Sugimoto, D. 1983, \JournalTitle{Astronomy and
  Astrophysics}, 126, 207

\bibitem[{Lecoanet {et~al.}(2016)Lecoanet, Schwab, Quataert, Bildsten, Timmes,
  Burns, Vasil, Oishi, \& Brown}]{lecoanet2016}
Lecoanet, D., Schwab, J., Quataert, E., {et~al.} 2016,
  \href{http://dx.doi.org/10.3847/0004-637X/832/1/71}{\JournalTitle{The
  Astrophysical Journal}, 832, 71}

\bibitem[{Li {et~al.}(2003)Li, Filippenko, Chornock, Berger, Berlind, Calkins,
  Challis, Fassnacht, Jha, Kirshner, Matheson, Sargent, Simcoe, Smith, \&
  Squires}]{Li:2003wja}
Li, W., Filippenko, A.~V., Chornock, R., {et~al.} 2003,
  \href{http://dx.doi.org/10.1086/374200}{\JournalTitle{Publications of the
  Astronomical Society of the Pacific}, 115, 453}

\bibitem[{Livio \& Mazzali(2018)}]{Livio:2018rue}
Livio, M., \& Mazzali, P. 2018,
  \href{http://dx.doi.org/10.1016/j.physrep.2018.02.002}{\JournalTitle{Physics
  Reports}, 736, 1}

\bibitem[{Lyman {et~al.}(2013)Lyman, James, Perets, Anderson, {Gal-Yam},
  Mazzali, \& Percival}]{Lyman:2013drq}
Lyman, J.~D., James, P.~A., Perets, H.~B., {et~al.} 2013,
  \href{http://dx.doi.org/10.1093/mnras/stt1038}{\JournalTitle{Monthly Notices
  of the Royal Astronomical Society}, 434, 527}

\bibitem[{Magee {et~al.}(2019)Magee, Sim, Kotak, Maguire, \&
  Boyle}]{Magee:2018aui}
Magee, M.~R., Sim, S.~A., Kotak, R., Maguire, K., \& Boyle, A. 2019,
  \href{http://dx.doi.org/10.1051/0004-6361/201834420}{\JournalTitle{Astronomy
  and Astrophysics}, 622, A102}

\bibitem[{Magee {et~al.}(2016)Magee, Kotak, Sim, Kromer, Rabinowitz, Smartt,
  Baltay, Campbell, Chen, Fink, {Gal-Yam}, Galbany, Hillebrandt, Inserra,
  Kankare, Le~Guillou, Lyman, Maguire, Pakmor, R{\"o}pke, Ruiter, Seitenzahl,
  Sullivan, Valenti, \& Young}]{Magee:2016vnu}
Magee, M.~R., Kotak, R., Sim, S.~A., {et~al.} 2016,
  \href{http://dx.doi.org/10.1051/0004-6361/201528036}{\JournalTitle{Astronomy
  and Astrophysics}, 589, A89}

\bibitem[{Maoz \& Badenes(2010)}]{Maoz:2010pz}
Maoz, D., \& Badenes, C. 2010,
  \href{http://dx.doi.org/10.1111/j.1365-2966.2010.16988.x}{\JournalTitle{Monthly
  Notices of the Royal Astronomical Society}, 407, 1314}

\bibitem[{Maoz {et~al.}(2014)Maoz, Mannucci, \& Nelemans}]{Maoz:2013hna}
Maoz, D., Mannucci, F., \& Nelemans, G. 2014,
  \href{http://dx.doi.org/10.1146/annurev-astro-082812-141031}{\JournalTitle{Annual
  Review of Astronomy and Astrophysics}, 52, 107}

\bibitem[{Marquardt {et~al.}(2015)Marquardt, Sim, Ruiter, Seitenzahl, Ohlmann,
  Kromer, Pakmor, \& R{\"o}pke}]{marquardt2015}
Marquardt, K.~S., Sim, S.~A., Ruiter, A.~J., {et~al.} 2015,
  \href{http://dx.doi.org/10.1051/0004-6361/201525761}{\JournalTitle{Astronomy
  and Astrophysics}, 580, A118}

\bibitem[{Meng \& Podsiadlowski(2014)}]{Meng:2014qta}
Meng, X., \& Podsiadlowski, P. 2014,
  \href{http://dx.doi.org/10.1088/2041-8205/789/2/L45}{\JournalTitle{\textbackslash{}apj},
  789, L45}

\bibitem[{Nomoto(1982)}]{Nomoto:1982zz}
Nomoto, K. 1982, \href{http://dx.doi.org/10.1086/159682}{\JournalTitle{The
  Astrophysical Journal}, 253, 798}

\bibitem[{Nomoto \& Kondo(1991)}]{nomoto1991}
Nomoto, K., \& Kondo, Y. 1991,
  \href{http://dx.doi.org/10.1086/185922}{\JournalTitle{The Astrophysical
  Journal}, 367, L19}

\bibitem[{Owocki {et~al.}(2004)Owocki, Gayley, \& Shaviv}]{Owocki:2004zz}
Owocki, S.~P., Gayley, K.~G., \& Shaviv, N.~J. 2004,
  \href{http://dx.doi.org/10.1086/424910}{\JournalTitle{The Astrophysical
  Journal}, 616, 525}

\bibitem[{Paxton {et~al.}(2011)Paxton, Bildsten, Dotter, Herwig, Lesaffre, \&
  Timmes}]{Paxton:2010ji}
Paxton, B., Bildsten, L., Dotter, A., {et~al.} 2011,
  \href{http://dx.doi.org/10.1088/0067-0049/192/1/3}{\JournalTitle{The
  Astrophysical Journal Supplement Series}, 192, 3}

\bibitem[{Paxton {et~al.}(2013)Paxton, Cantiello, Arras, Bildsten, Brown,
  Dotter, Mankovich, Montgomery, Stello, Timmes, \& Townsend}]{Paxton:2013pj}
Paxton, B., Cantiello, M., Arras, P., {et~al.} 2013,
  \href{http://dx.doi.org/10.1088/0067-0049/208/1/4}{\JournalTitle{The
  Astrophysical Journal Supplement Series}, 208, 4}

\bibitem[{Paxton {et~al.}(2015)Paxton, Marchant, Schwab, Bauer, Bildsten,
  Cantiello, Dessart, Farmer, Hu, Langer, Townsend, Townsley, \&
  Timmes}]{Paxton:2015jva}
Paxton, B., Marchant, P., Schwab, J., {et~al.} 2015,
  \href{http://dx.doi.org/10.1088/0067-0049/220/1/15}{\JournalTitle{The
  Astrophysical Journal Supplement Series}, 220, 15}

\bibitem[{Paxton {et~al.}(2018)Paxton, Schwab, Bauer, Bildsten, Blinnikov,
  Duffell, Farmer, Goldberg, Marchant, Sorokina, Thoul, Townsend, \&
  Timmes}]{Paxton:2017eie}
Paxton, B., Schwab, J., Bauer, E.~B., {et~al.} 2018,
  \href{http://dx.doi.org/10.3847/1538-4365/aaa5a8}{\JournalTitle{The
  Astrophysical Journal Supplement Series}, 234, 34}

\bibitem[{Podsiadlowski {et~al.}(2004)Podsiadlowski, Langer, Poelarends,
  Rappaport, Heger, \& Pfahl}]{Podsiadlowski:2003py}
Podsiadlowski, P., Langer, N., Poelarends, A. J.~T., {et~al.} 2004,
  \href{http://dx.doi.org/10.1086/421713}{\JournalTitle{The Astrophysical
  Journal}, 612, 1044}

\bibitem[{Poelarends(2007)}]{poelarends2007}
Poelarends, A. J.~T. 2007, PhD thesis

\bibitem[{Poelarends {et~al.}(2008)Poelarends, Herwig, Langer, \&
  Heger}]{Poelarends:2007ip}
Poelarends, A. J.~T., Herwig, F., Langer, N., \& Heger, A. 2008,
  \href{http://dx.doi.org/10.1086/520872}{\JournalTitle{The Astrophysical
  Journal}, 675, 614}

\bibitem[{Poelarends {et~al.}(2017)Poelarends, Wurtz, Tarka, Cole~Adams, \&
  Hills}]{Poelarends:2017dua}
Poelarends, A. J.~T., Wurtz, S., Tarka, J., Cole~Adams, L., \& Hills, S.~T.
  2017, \href{http://dx.doi.org/10.3847/1538-4357/aa988a}{\JournalTitle{The
  Astrophysical Journal}, 850, 197}

\bibitem[{Rose(1969)}]{rose1969}
Rose, W.~K. 1969, \href{http://dx.doi.org/10.1086/149885}{\JournalTitle{The
  Astrophysical Journal}, 155, 491}

\bibitem[{{Ruiz-Lapuente} {et~al.}(1993){Ruiz-Lapuente}, Jeffery, Challis,
  Filippenko, Kirshner, Ho, Schmidt, S{\'a}nchez, \& Canal}]{ruiz-lapuente1993}
{Ruiz-Lapuente}, P., Jeffery, D.~J., Challis, P.~M., {et~al.} 1993,
  \href{http://dx.doi.org/10.1038/365728a0}{\JournalTitle{Nature}, 365, 728}

\bibitem[{Sana {et~al.}(2012)Sana, {de Mink}, {de Koter}, Langer, Evans,
  Gieles, Gosset, Izzard, Le~Bouquin, \& Schneider}]{Sana:2012px}
Sana, H., {de Mink}, S.~E., {de Koter}, A., {et~al.} 2012,
  \href{http://dx.doi.org/10.1126/science.1223344}{\JournalTitle{Science}, 337,
  444}

\bibitem[{Sato {et~al.}(2015)Sato, Nakasato, Tanikawa, Nomoto, Maeda, \&
  Hachisu}]{Sato:2015spa}
Sato, Y., Nakasato, N., Tanikawa, A., {et~al.} 2015,
  \href{http://dx.doi.org/10.1088/0004-637X/807/1/105}{\JournalTitle{The
  Astrophysical Journal}, 807, 105}

\bibitem[{Schwab {et~al.}(2017)Schwab, Bildsten, \& Quataert}]{Schwab:2017epw}
Schwab, J., Bildsten, L., \& Quataert, E. 2017,
  \href{http://dx.doi.org/10.1093/mnras/stx2169}{\JournalTitle{Monthly Notices
  of the Royal Astronomical Society}, 472, 3390}

\bibitem[{Schwab {et~al.}(2015)Schwab, Quataert, \& Bildsten}]{Schwab:2015bma}
Schwab, J., Quataert, E., \& Bildsten, L. 2015,
  \href{http://dx.doi.org/10.1093/mnras/stv1804}{\JournalTitle{Monthly Notices
  of the Royal Astronomical Society}, 453, 1910}

\bibitem[{Schwab \& Rocha(2019)}]{Schwab:2018cnb}
Schwab, J., \& Rocha, K.~A. 2019,
  \href{http://dx.doi.org/10.3847/1538-4357/aaffdc}{\JournalTitle{The
  Astrophysical Journal}, 872, 131}

\bibitem[{{Seitenzahl} \& {Townsley}(2017)}]{Seitenzahl2017}
{Seitenzahl}, I.~R., \& {Townsley}, D.~M. 2017, {Nucleosynthesis in
  Thermonuclear Supernovae}, 1955

\bibitem[{Siess(2006)}]{siess2006}
Siess, L. 2006,
  \href{http://dx.doi.org/10.1051/0004-6361:20053043}{\JournalTitle{Astronomy
  and Astrophysics}, 448, 717}

\bibitem[{{Smith} \& {Owocki}(2006)}]{Smith2006}
{Smith}, N., \& {Owocki}, S.~P. 2006,
  \href{http://dx.doi.org/10.1086/506523}{\JournalTitle{\apjl}, 645, L45}

\bibitem[{Soker(2019)}]{soker2019}
Soker, N. 2019, \JournalTitle{arXiv e-prints}, 1912, arXiv:1912.01550

\bibitem[{Suzuki {et~al.}(2016)Suzuki, Toki, \& Nomoto}]{Suzuki:2015iry}
Suzuki, T., Toki, H., \& Nomoto, K. 2016,
  \href{http://dx.doi.org/10.3847/0004-637X/817/2/163}{\JournalTitle{The
  Astrophysical Journal}, 817, 163}

\bibitem[{Takahashi {et~al.}(2013)Takahashi, Yoshida, \&
  Umeda}]{Takahashi:2013ena}
Takahashi, K., Yoshida, T., \& Umeda, H. 2013,
  \href{http://dx.doi.org/10.1088/0004-637X/771/1/28}{\JournalTitle{The
  Astrophysical Journal}, 771, 28}

\bibitem[{Taubenberger(2017)}]{Taubenberger:2017hoo}
Taubenberger, S. 2017,
  \href{http://dx.doi.org/10.1007/978-3-319-21846-5_37}{\JournalTitle{Handbook
  of Supernovae}, 317}

\bibitem[{Tsebrenko \& Soker(2015)}]{Tsebrenko:2014kea}
Tsebrenko, D., \& Soker, N. 2015,
  \href{http://dx.doi.org/10.1093/mnras/stu2567}{\JournalTitle{Monthly Notices
  of the Royal Astronomical Society}, 447, 2568}

\bibitem[{van Kerkwijk {et~al.}(2010)van Kerkwijk, Chang, \&
  Justham}]{vanKerkwijk:2010he}
van Kerkwijk, M.~H., Chang, P., \& Justham, S. 2010,
  \href{http://dx.doi.org/10.1088/2041-8205/722/2/L157}{\JournalTitle{The
  Astrophysical Journal}, 722, L157}

\bibitem[{Vink(2017)}]{Vink:2017ujd}
Vink, J.~S. 2017,
  \href{http://dx.doi.org/10.1051/0004-6361/201731902}{\JournalTitle{Astronomy
  and Astrophysics}, 607, L8}

\bibitem[{Waldman \& Barkat(2006)}]{waldman2006a}
Waldman, R., \& Barkat, Z. 2006, \JournalTitle{Ph.D. Thesis}

\bibitem[{{Waldman} {et~al.}(2008){Waldman}, {Yungelson}, \&
  {Barkat}}]{waldman2008}
{Waldman}, R., {Yungelson}, L.~R., \& {Barkat}, Z. 2008, in Astronomical
  Society of the Pacific Conference Series, Vol. 391, Hydrogen-Deficient Stars,
  ed. A.~{Werner} \& T.~{Rauch}, 359

\bibitem[{Wang \& Han(2012)}]{Wang:2012za}
Wang, B., \& Han, Z. 2012,
  \href{http://dx.doi.org/10.1016/j.newar.2012.04.001}{\JournalTitle{New
  Astronomy Reviews}, 56, 122}

\bibitem[{Wheeler(1978)}]{wheeler1978}
Wheeler, J.~C. 1978, \href{http://dx.doi.org/10.1086/156484}{\JournalTitle{The
  Astrophysical Journal}, 225, 212}

\bibitem[{Woosley(2019)}]{Woosley:2019sdf}
Woosley, S.~E. 2019,
  \href{http://dx.doi.org/10.3847/1538-4357/ab1b41}{\JournalTitle{The
  Astrophysical Journal}, 878, 49}

\bibitem[{Yoon(2017)}]{Yoon:2017dme}
Yoon, S.-C. 2017,
  \href{http://dx.doi.org/10.1093/mnras/stx1496}{\JournalTitle{Monthly Notices
  of the Royal Astronomical Society}, 470, 3970}

\end{thebibliography}
